\documentclass[amsmath,amssymb,aps,prb,twocolumn,footinbib]{revtex4-1}

\usepackage{subfigure}
\usepackage{enumitem}
\usepackage[colorlinks=true,linkcolor=blue,urlcolor=blue,citecolor=blue]{hyperref}
\usepackage{graphicx}
\usepackage{dcolumn}
\usepackage{bm}
\usepackage{mathtools}

\newcommand{\red}{}

\begin{document}

\title{Multifractality at the Weyl semimetal-diffusive metal transition for generic disorder}

\author{Eric Brillaux}
\author{David Carpentier}%
\author{Andrei A. Fedorenko}%
\affiliation{ \mbox{Univ Lyon, ENS de Lyon, Univ Claude Bernard, CNRS, Laboratoire de Physique, F-69342 Lyon, France}}

\date{\today}

\begin{abstract}

A Weyl semimetal is a three dimensional topological gapless phase.
In the presence of strong enough disorder it undergoes a quantum transition towards a diffusive metal phase whose universality class depends on the range of disorder correlations.
Similar to other quantum transitions driven by disorder, the critical wave functions
at the semimetal-diffusive metal transition exhibit multifractality.
Using renormalization group methods we study the corresponding multifractal spectrum
as a function of the range of disorder correlations for generic disorder
including random scalar and vector potentials. We also discuss the relation between
the geometric fluctuations of critical wave functions and the broad distribution of
the local density of states (DOS) at the transition. We derive a new scaling relation
for the typical  local DOS and  argue that it holds for other
disorder-driven transitions in which both the average and typical  local DOS vanish on one side
of the transition. As an illustration we apply it to the recently discussed unconventional quantum transition in disordered semiconductors with power-law dispersion relation near the band edge.

\end{abstract}

\maketitle


\section{\label{sec:intro}Introduction}

Whereas our basic understanding of solids is based on a description as a perfectly regular lattice of atoms,
real materials do not meet this requirement. The presence of disorder such as lattice defects or impurities can obscure properties of ideal solids, or even lead to new quantum phenomena such as the Anderson localization.\cite{Abrahams:2010} Recently a new type of disorder-driven  quantum phase transition was discovered in three dimensional relativistic semimetals.~\cite{Syzranov:2018} In these topological materials, several bands cross linearly at isolated points in the Brillouin zone: two bands in Weyl  semimetals~\cite{Xu:2015a,Xu:2015b} and four bands in Dirac semimetals.\cite{Liu:2014,Neupane:2014,Borisenko:2014} Many aspects of relativistic semimetals
were discussed in the past,\cite{Balents:2011,Wan:2011,Wang:2012} but the compounds that host them were identified experimentally only recently.\cite{Lv:2015,Xu2015}
These materials immediately attracted a lot of attention because the relativistic nature of low energy excitations lead to peculiar properties, such as the anomalous quantum
Hall effect,\cite{Yang2011} the chiral anomaly,\cite{Yan2016,Burkov2016,Rao2016,Armitage2018}
and the related negative magnetoresitance.\cite{Son:2013,Burkov2015,Liang2018}

Disorder also leads to remarkable properties: while weak disorder is irrelevant for
relativistic electrons in three dimensions, a strong enough disorder drives
the semimetal towards a diffusive metal.
The average DOS at the nodal point, which plays the role of an order parameter,
becomes non-zero above the critical disorder strength $\Delta^*$ and behaves as
$\bar{\rho}(0) \sim (\Delta-\Delta^*)^{\beta}$, while the correlation length diverges as
$\xi \sim (\Delta-\Delta^*)^{-\nu}$.\cite{Sbierski:2014, Sbierski:2016,Fradkin:1986,Roy:2016b,Goswami:2011,Hosur:2012,Ominato:2014,Chen:2015,Altland:2015:2016}
This disorder-driven transition has been intensively studied using both numerical
simulations\cite{Kobayashi:2014,Sbierski:2015,Liu:2016,Bera:2016,Fu:2017,Sbierski:2017}
and analytical methods.\cite{Syzranov:2015b,Roy:2014,Louvet:2016,Balog:2018,Sbierski:2019,Klier:2019}
The effects of rare events have also been much debated.\cite{Holder:2017,Gurarie:2017} Rare fluctuations of disorder potential might create an exponentially small but finite DOS in the semimetal phase, thus {\red rounding the sharp} transition;\cite{Nandkishore:2014,Pixley:2016,Pixley:2016c,Wilson:2017,Wilson:2018,Ziegler:2018} however, the probability
of such fluctuations turns out to be extremely small.\cite{Buchhold:2018,Buchhold:2018-2}

Besides the average DOS, other indicators help pinpoint
the critical behavior. In particular the critical wave functions exhibit a multifractal behavior at the semimetal-diffusive metal transition.~\cite{Syzranov2016,Louvet:2016}
The inverse participation ratios $ P_q = \int \mathrm{d}r \overline{ {|\psi(\bm{r})|^{2q}}}$
averaged over disorder scale with the system size $L$ as $ P_q \sim L^{-\tilde{\tau}_q}$, where the multifractal spectrum exponent $\tilde{\tau}_q$ depends non-linearly on $q$. The multifractal spectrum encodes much more information
about the transition than just the behavior of the average DOS.
Remarkably the multifractal spectrum at the semimetal-diffusive
metal transition differs from any Wigner-Dyson universality class relevant for the Anderson localization.\cite{Wegner1987}
Using the $\varepsilon=d-2$ expansion we find that the non-linearity of the multifractal exponent
is of order $\varepsilon^2$ for the semimetal-diffusive metal transition\cite{Syzranov2016,Louvet:2016} while for the unitary and orthogonal classes it is of order $\varepsilon^{1/2}$ and $\varepsilon$, respectively.\cite{Evers:2008}

The aforementioned studies disregard defects that are correlated over large distances. However, the presence of linear dislocations, or {\red grain boundaries}
are known to generate long-range correlations in the disorder distribution.
Introducing disorder correlations is also widely used in numerical simulations
to decouple different Weyl cones by suppressing inter-valley scattering.
Power-law correlations decaying with the distance $r$ as
$1/r^a$ may drive a continuous transition to a new universality class,\cite{Weinrib1983,Fedorenko2012} and
numerical simulations suggest that they modify the critical exponents at the Anderson localization transition.\cite{Croy2012}
The effect of disorder  correlations on the semimetal-diffusive metal transition was
investigated in Ref.~\onlinecite{Louvet:2017}. It was found that
for $a<2$ the semimetal phase is always unstable while for $2<a<a_c\approx 2.8 $ disorder drives the
transition to a new universality class, whose critical exponents depend on~$a$. We thus expect multifractality to be affected by disorder correlations.

In this paper we investigate the multifractal spectrum at the semimetal-diffusive metal transition in the presence of a generic type of disorder. We study critical fluctuations of the DOS  and compute both the  average local DOS and the  typical local DOS at the transition.
The paper is organized as follows.
Section~\mbox{\ref{sec:model}} introduces the model of Weyl fermions in the presence of correlated scalar disorder. In Sec.~\mbox{\ref{sec:multifractality}}
we compare the multifractal spectra for the Anderson localization transition and for the semimetal-diffusive metal transition, and show that the way the moments of the DOS distribution behave make it possible to distinguish between different phases. In Sec.~\ref{sec:TDOS} we derive the scaling relations for the exponents $\beta$ and $\beta_{\mathrm{typ}}$ which describe the critical behavior of the average and typical local DOS, respectively. In Sec.~\mbox{\ref{sec:unconventional}} we apply our scaling relations to the unconventional disorder-driven transition in  semiconductors with power-law dispersion relation near the band edge. Then in Sec.~\mbox{\ref{sec:RGflow}} we present the renormalization
group picture, and derive the multifractal spectrum to two-loop order. In Sec.~\mbox{\ref{sec:vectorial}} we generalize our approach to vector potential disorder,
and show that this type of disorder does not affect criticality even in the presence of long-range correlations. Section~\mbox{\ref{sec:discussion}} summarizes our findings.

\section{\label{sec:model} Model}

We consider a single Weyl node subject to scalar quenched disorder.
Though Weyl nodes always come in pairs of opposite chiralities,\cite{Nielsen:1981} we may neglect inter-node
scattering provided the correlation length $\xi_d$ of disorder is much greater than the inverse
of the separation $b$ of the nodes in the Brillouin zone.\cite{Altland:2015:2016}
We assume that Coulomb repulsion between electrons is negligible and that the node lies at
the Fermi energy $\varepsilon_F = 0$.

The low energy Hamiltonian which describes non-interacting three dimensional Weyl fermions moving in the scalar,
time-independent potential $V(\bm{r})$ created by impurities reads
\begin{equation} \label{eq:Ham1}
H(\bm{r}) = -i v_F \, \bm{\sigma} \! \cdot \! \bm{\partial} + V(\bm{r}) \mathbb{I},
\end{equation}
where $v_F$ is the Fermi velocity (from now on set to one) and
$\bm{\sigma}=(\sigma_1,\sigma_2,\sigma_3)$ are the Pauli matrices, {\red and $\mathbb{I}$ is the identity matrix.}
In order to use dimensional regularization we define the Hamiltonian~(\ref{eq:Ham1}) in arbitrary dimension $d$
by generalizing the Pauli matrices to a Clifford algebra satisfying the anticommutation relations:
$\gamma_i \gamma_j + \gamma_j \gamma_i = 2 \delta_{ij} \mathbb{I}$ ($i,j=1,...,d$).
{\red Since the fermions are non-interacting and the disorder potential is time independent,
it is convenient to write down the corresponding action at fixed Matsubara frequency $\omega$ as
\begin{equation}
\label{eq:action0}
S  =
	\int d^d r\,  \bar{\psi}(\mathbf{r})\left[ - i \bm{\gamma} \! \cdot \! \bm{\partial}  - i \omega     + V(\mathbf{r}) \mathbb{I} \right]   \psi(\mathbf{r}) ,
\end{equation}
where $\bar{\psi}$ and $\psi$ are two conjugate Weyl spinors.}
We assume that the distribution of disorder potential is translationally invariant, isotropic and
Gaussian with the mean value and variance given by
\begin{equation}
\overline{V(\bm{r})} = 0, \ \ \    \overline{V(\bm{r}) V(\bm{0})} = g(r),
\end{equation}
where the overbar indicates an average over many disorder realizations. The short-range correlation $g(r)$ is generally approximated by a Gaussian function whose width $\xi_d$ may be set to zero close enough to the transition where $\xi \gg \xi_d$,
so that $g(r) \sim \Delta_S \, \delta^d(\bm{r})$. {\red In numerical simulations on a lattice one usually chooses
$\xi_d \gg b^{-1}$ to suppress inter-node scattering and for small $b$ there exists
a wide range of scales at which the effective correlations can be approximated by a power-law. In addition,
the presence of extended defects in the form of linear dislocations or grain boundaries can lead to a power-law decay of the correlation,\cite{Fedorenko:2006}
$g(r) \sim \Delta_L r^{-a}$. Another possible source for power-law correlations is the presence of Coulomb impurities. However, in this case the system is unstable with respect to the formation of electron and hole puddles with localized states at zero energy.  This creates an algebraically finite DOS at zero chemical potential for arbitrary weak disorder,  which smears out the transition.\cite{Skinner2014}
}
Since the short-range correlations are ultimately generated by the renormalization flow,
we account for both short-range and long-range contributions and write in Fourier space\cite{Louvet:2017}
\begin{equation}
g(k) = \Delta_S + \Delta_L k^{a-d}. \label{eq-dis-cor-range}
\end{equation}

To average over disorder we use the replica trick.  {\red Introducing
 $n$ replicas of the original system and averaging over the potential distribution,
we arrive at}
\begin{multline}
\label{eq:effective_action}
S_{\text{eff}} [\bar{\psi}_{\alpha},\psi_{\alpha}] =  \int_{k} \bar{\psi}_{\alpha} (-\bm{k})
 ( \bm{\gamma} \! \cdot \! \bm{k} - i\omega )
\psi_{\alpha} (\bm{k}) \\
-\frac{1}{2} \int_{k_i } (\Delta_{S} + \Delta_{L} |\bm{k_1} + \bm{k_2}|^{a-d}) \bar{\psi}_{\alpha}
(\bm{k_1}) \psi_{\alpha} (\bm{k_2}) \\
\bar{\psi}_{\beta} (\bm{k_3})\psi_{\beta} (\bm{k_1}+\bm{k_2}-\bm{k_3}),
\end{multline}
where summation over repeated replica indices $\alpha$ and $\beta$ is assumed.
The properties of the original system averaged over disorder are recovered in the limit of $n \rightarrow 0$.

\section{Multifractal spectrum}\label{sec:multifractality}

The notion of multifractality has turned out to be useful in many physical problems
ranging from turbulence\cite{Parisi:1985} to disordered classical spin
systems\cite{Duplantier:1991} and the Anderson localization transition.\cite{Evers:2008}
Similar to the latter example,  the critical wave functions at the semimetal-diffusive metal transition exhibit multifractality so that their geometrical properties can be described  by the multifractal  spectrum. As we show below, this spectrum also encodes  the scaling behavior of the whole distribution of the local DOS at the transition. It is instructive to compare the scaling properties of the wave functions and local DOS fluctuations at the Anderson localization and the semimetal-diffusive metal transition and highlight the similarities and differences between these two disorder-driven quantum transitions.

Let us first recall the major results on the Anderson localization transition. The statistical properties of wave functions close to the mobility edge $\omega_c$  can be described using either the participation ratio or the inverse participation ratio, depending on the phase from which we approach criticality.

In the region of localized states ($\omega<\omega_c$) it is convenient to introduce
the inverse participation ratio (IPR)\cite{Wegner1980}
\begin{equation}
 {P}_q (\omega) = \frac{\int \mathrm{d}^d r\overline{\sum_i|\psi_i(\mathbf{r})|^{2q}\delta(\omega-\omega_i)}
 }{\int \mathrm{d}^d r \overline{\rho(\mathbf{r},\omega)}},
 \label{eq:IPR-1}
\end{equation}
where $\psi_i(\mathbf{r})$ is an eigenstate with energy $\omega_i$ and
$\rho(\mathbf{r},\omega)=\sum_i|\psi_i(\mathbf{r})|^{2}\delta(\omega-\omega_i)$
is the local DOS.
The IPR (\ref{eq:IPR-1})
gives the $q$-moment of the inverse volume spanned by the localized wave function. It vanishes in the region of extended states ($\omega>\omega_c$), but decays in a power-law fashion in the region of localised states ($\omega<\omega_c$) as ${P}_q (\omega)\sim (\omega_c-\omega)^{\pi_q} $ in the thermodynamic limit. For a finite system precisely at the mobility edge
the IPR scales with the size of the system $L$ as ${P}_q (\omega_c) \sim L^{-\tilde{\tau}_q}$ where $\tilde{\tau}_q \nu = \pi_q$ and $\nu$ is the critical exponent for the correlation length, $\xi \sim |\omega-\omega_c|^{-\nu}$.

In the region of extended states it is more convenient to consider
the participation ratio (PR)
$p_q(\omega)$ defined by
\begin{equation}
 \frac1{p_q (\omega)L^{d(q-1)}} = \frac{  \int \mathrm{d}^dr\overline{ \rho^q(\mathbf{r},\omega)}}{[  \int \mathrm{d}^dr \overline{\rho(\mathbf{r},\omega)}]^q},
 \label{eq:PR}
\end{equation}
which gives the $q$-moment of the fraction of sites occupied by the wave function. Contrary to the IPR~(\ref{eq:IPR-1}) the PR~(\ref{eq:PR}) vanishes in the region of localized states ($\omega<\omega_c$), but decays as ${p}_q (\omega)\sim (\omega-\omega_c)^{\mu_q} $ for
$\omega>\omega_c$ in the thermodynamic limit. The r.h.s. of both equations
(\ref{eq:IPR-1}) and (\ref{eq:PR}) scale identically with the size of the system $L$ at the transition, which imposes the relation $\pi_q = d\nu(q-1)-\mu_q$ {\red between the two exponents.\cite{Wegner1980}}

Introducing the scaling dimension $x^*_q$ of the $q$-moment of the local DOS, such that   $\overline{\rho^q(\mathbf{r},\omega_c)}\sim L^{-x^*_q}$, and using Eq.~(\ref{eq:PR}) we arrive at
\begin{equation}
\tilde{\tau}_q = d(q-1) + \tilde{\Delta}_q,
 \label{eq:tau-def}
\end{equation}
where we split the multifractal spectrum exponent into the normal part $d(q-1)$ corresponding to the metallic phase and the anomalous dimension
$\tilde{\Delta}_q=x^*_q-q x^*_1$. The anomalous dimension
$\tilde{\Delta}_q$ gives the scaling behavior of the normalized $q$-moment of the local DOS,
\begin{equation}
\overline{\tilde{\rho}^q } \sim L^{-\tilde{\Delta}_q}, \ \ \ \  \tilde{\rho} = \frac{\rho}{\overline{\rho}}.
 \label{eq:tau-def2}
\end{equation}
The Legendre transform of $\tilde{\tau}_q$
\begin{equation}
\tilde{f}(\alpha) = \alpha q - \tilde{\tau}_q, \ \ \ \alpha(q) = \mathrm{d}\tilde{\tau}_q/\mathrm{d}q,
\end{equation}
is known as the singularity spectrum\cite{Evers:2008}
and gives the fractal dimension of the manifold spanning the points with the wave function intensity
$|\psi(\mathbf{r})|^2 \sim L^{-\alpha}$, i.e. the volume of this manifold scales with the system size as
$L^{\tilde{f}(\alpha)}$.\cite{Halsey1987}

The above picture can be contrasted with that for the semimetal-diffusive metal transition. In the latter case the transition occurs at $\omega=0$ and is driven by the strength of disorder $\Delta$. For instance, the correlation length diverges as $\xi \sim |\Delta-\Delta^*|^{-\nu}$.
The main difference, however, lies in the behavior of the average DOS. While at the Anderson transition
the average local DOS varies smoothly without vanishing across the critical point, in the case of the
semimetal - diffusive metal transition it behaves as
\begin{equation}
\bar{\rho}(\Delta) \sim (\Delta - \Delta^*)^\beta
 \label{eq:rho-12}
\end{equation}
in the metal phase ($\Delta > \Delta^*$), and
vanishes in the semimetal phase ($\Delta < \Delta^*$).
The exponent $\beta$ describing the scaling behavior of the average local DOS is related to the dynamic critical exponent $z$ by\cite{Kobayashi:2014}
\begin{equation}
\beta = \nu(d-z).
 \label{eq:rho-13}
\end{equation}
As a consequence the IPR $P_q(\Delta, \omega=0)$ vanishes everywhere in the thermodynamic limit, but its finite size scaling at the critical point $\Delta=\Delta^*$ reads
\begin{equation}
{P}_q (\Delta=\Delta^*,\omega=0) \sim L^{-\tilde{\tau}_q},
 \label{eq:Pq-scaling-crit}
\end{equation}
with the exponent $\tilde{\tau}_q$ given by Eq.~(\ref{eq:tau-def}). Remarkably, the $q$-moment
of the fraction of sites occupied by the wave function behaves as
\begin{equation}
{p}_q (\Delta,\omega=0) \sim \begin{cases}
    0, & \text{$\Delta<\Delta^*$}.\\
    (\Delta-\Delta^*)^{-\tilde{\Delta}_q/\nu}, & \text{$\Delta>\Delta^*$},
  \end{cases}
   \label{eq:tau-def3}
\end{equation}
with $\tilde{\Delta}_q<0$ for $q>1$. Thus, the PR ${p}_q (\Delta,\omega=0)$ for $q>1$ can
play the role of an order parameter.

{\red It was recently argued that the presence of rare large regions with strong disorder potential can create a finite DOS at zero energy even for weak disorder, thus, rendering the transition to be avoided.\cite{Nandkishore:2014,Pixley:2016}
This would introduce a new length scale above which the wave function is not multifractal, or at least not with the same multifractal spectrum. However, our results on the multifractal spectrum still apply below this crossover length scale.

While the exponentially small DOS has been detected numerically, the theoretical picture of this phenomenon is still controversial. Two scenarios have been proposed. In the first scenario, a zero energy state is created by an optimal fluctuation of disorder potential (instanton).\cite{Nandkishore:2014} In three dimensions, in order to (quasi)localize a relativistic particle, the potential of the well has to decay with the distance $r$ to the center as a power-law $1/r^4$.  However, as was shown in Ref.~\onlinecite{Buchhold:2018} by expanding and integrating out the Gaussian fluctuations around this instanton solution, the prefactor in front of the exponentially small DOS vanishes at zero energy. In the second scenario the finite DOS is generated  by resonances between two different rare regions with strong disorder,\cite{Ziegler:2018} but as was argued in Ref.~\onlinecite{Buchhold:2018-2} these resonances cannot create states exactly at zero energy.

If the disorder potential is a random Gaussian field, the probability to have the above large regions of strong disorder potential is inversely proportional to the exponential of this potential squared and integrated over space. It is clear that in both scenarios this probability is much smaller than that in the usual Lifshitz tail problems where the integrals are taken over exponentially and not power-law decaying  instanton solutions.\cite{Falco:2017} This implies that the corresponding crossover length scale has to be very large.
Indeed, the critical behavior is accessible in numerical simulation despite the presence of rare events.\cite{Pixley:2016,Pixley:2016c}
Moreover, we argue that the order parameter~(\ref{eq:tau-def3}) can be better used to characterize the transition in this case. This is because the (quasi)localized  zero energy states decay as $1/r^2$ in three dimensions, and thus, cannot create a finite ${p}_q (\Delta,\omega=0)$ for $\Delta<\Delta^*$ since the fraction of sites occupied by a normalizable wavefunction vanishes in the thermodynamic limit.}


\section{Typical vs average DOS} \label{sec:TDOS}

The local DOS has a broad distribution at the Anderson transition so that typical and average local DOS behave quite differently.
The average DOS $\rho(\omega) = \overline{\rho(\mathbf{r},\omega)}$ varies smoothly around the critical point and does not exhibit any qualitative change upon localization. The typical DOS
$\rho_{\mathrm{typ}}(\omega) = \exp \overline{\ln \rho(\mathbf{r},\omega)}$ is finite in the delocalized phase, decreases when approaching the transition, and vanishes in the localized phase.
The reason is that upon localization the local spectrum changes from a continuous to an essentially discrete one. Since the local DOS directly probes
the local amplitudes of wave functions the typical value of the local DOS is zero in the last case.

A similar argument applies to the semimetal-diffusive metal transition where one also expects a broad
distribution of local DOS and different behaviors for the average and typical DOS.\cite{Balog:2018} Contrary to the Anderson transition, both typical and average DOS vanish in the semimetal phase but with different exponents, in particular
\begin{eqnarray}
&&\rho_{\mathrm{typ}}(\Delta) \sim (\Delta-\Delta^*)^{\beta_\mathrm{typ}}, \label{eq-beta-typ-1}
\end{eqnarray}
where $\beta_\mathrm{typ}$ differs from the average DOS exponent $\beta$~(\ref{eq:rho-13}).

To determine the exponent $\beta_\mathrm{typ}$, let us consider the distribution $\mathcal{P}(\tilde{\rho},L)$ of the normalized local DOS $\tilde{\rho} = \rho/\overline{\rho}$  in a finite size system.
Its moments follow the scaling law~\eqref{eq:tau-def2} and read
\begin{eqnarray}
&& \overline{\tilde{\rho}^q} = \int_0^\infty d \tilde{\rho}\,  \tilde{\rho}^q \mathcal{P}(\tilde{\rho},L) = c_q L^{-\tilde{\Delta}_q},
\end{eqnarray}
where $c_q$ depends weakly on $L$. We now change variable from $\tilde{\rho}$ to $\alpha$ such that\cite{Janssen1998}
$\tilde{\rho} = L^{-\alpha}$  and  $\mathcal{P}(\tilde{\rho},L)d\tilde{\rho} = \tilde{\mathcal{P}} (\alpha,L) d\alpha$.
We arrive at
\begin{eqnarray} \label{eq:Laplace1}
&& \overline{\tilde{\rho}^q} = \int d \alpha \exp \left[ \ln L ( \tilde{g}(\alpha) - \alpha q )  \right]
\end{eqnarray}
with  $\tilde{g}(\alpha) = \ln \tilde{P} / \ln L$. Noticing the large prefactor $\ln L$ in the exponential of Eq.~(\ref{eq:Laplace1}), we apply the steepest descent method and find that $\tilde{g}(\alpha)$ is the Legendre transform of the anomalous dimension
\begin{equation}
\tilde{g}(\alpha(q)) = \alpha q - \tilde{\Delta}_q, \ \ \  \alpha(q) = \mathrm{d}\tilde{\Delta}_q/\mathrm{d}q,
\end{equation}
and can be expressed in terms of the singularity spectrum as $\tilde{g}(\alpha) = \tilde{f} (\alpha+d)-d$. This function peaks at $\alpha=\alpha_0-d$, where $\alpha_0$ is the position of the peak of the singularity spectrum.
It gives the most probable scaling exponent which describes the scaling behavior of the typical normalized
local DOS
\begin{equation}
\tilde{\rho}_{\text{typ}} = \exp \int  d \tilde{\rho}  \,  \mathcal{P}(\tilde{\rho},L)  \ln \tilde{\rho } \sim L^{d-\alpha_0}.
\end{equation}
From the scaling dimension of $\tilde{\rho}_{\text{typ}}$ we deduce that near the critical point on the metal side of the transition,
\begin{equation}
\tilde{\rho}_{\text{typ}}(\Delta) \sim (\Delta-\Delta^*)^{\nu(\alpha_0-d)}.
\end{equation}
Using Eqs.~(\ref{eq:rho-12})-(\ref{eq:rho-13}) and $\tilde{\rho}_{\text{typ}} = {\rho}_{\text{typ}} / \bar{\rho} $  we find that the typical local DOS vanishes at the transition according to Eq.~(\ref{eq-beta-typ-1}) with the exponent
\begin{equation}
{\beta_\mathrm{typ}}= \nu(\alpha_0-z). \label{eq-scal-rel-2}
\end{equation}
We can compare this exponent with that for the typical DOS at the Anderson transition given by the scaling relation ${\beta_\mathrm{typ}}= \nu(\alpha_0-d)$.\cite{Janssen1998,Pixley2015}
It differs from Eq.~(\ref{eq-scal-rel-2}) due to the smooth, non-vanishing behavior of the average local DOS
around the localization point.

It turns out that numerical simulations
give large errors for the critical exponent $\nu$, so that it is useful to derive from
Eq.~(\ref{eq-scal-rel-2}) a scaling relation wherein $\nu$ is absent, as in
\begin{equation}
\frac{\beta_\mathrm{typ}}{\beta} =\frac{\alpha_0-z}{d-z}. \label{beta-typ-scaling}
\end{equation}
{\red For short-range (SR) correlated disorder,} the numerical simulations of Refs.~\onlinecite{Pixley2015,Pixley2016} give
$\beta^{{\rm SR}} = 1.4 \pm 0.2$, $\beta^{{\rm SR}}_{\mathrm{typ}} = 2.0 \pm 0.3$, and $z^{{\rm SR}} =1.46 \pm 0.05$. Using these values, we estimate the position of the singularity spectrum peak at
\begin{equation}
 \alpha_0^{{\rm SR}} = z^{{\rm SR}} + (d-z^{{\rm SR}}) \dfrac{\beta^{{\rm SR}}_{\mathrm{typ}}}{\beta^{{\rm SR}}} = 3.7 \pm 0.6.
 \label{eq:alpha-zero-num}
\end{equation}
In Sec.~\ref{sec:RGflow} we calculate the anomalous dimension $\tilde{\Delta}_q$ and the exponent $\alpha_0$ as a function of the disorder correlations of Eq.~\eqref{eq-dis-cor-range} to two-loop order, and compare our analytical prediction to Eq.~\eqref{eq:alpha-zero-num}.

{\red
The scaling relations \eqref{eq-beta-typ-1} and \eqref{eq-scal-rel-2} for the typical DOS exponents constitute one of the main results of this work.
Before we use these results to describe the Weyl semimetal-diffusive metal  transition
we would like to emphasize that they are more general and apply to other disorder-induced transitions. As an illustration we consider the unconventional transition in high-dimensional disordered semiconductors.}

\section{\label{sec:unconventional}
Unconventional transition in disordered semiconductors }

{\red
The relations for the critical exponents describing the typical and average DOS behavior
hold not only for the Weyl semimetal-diffusive metal transition, but also for other disorder-driven transitions, provided that the critical wave functions exhibit multifractality and both the typical and average DOS vanish on one side of the transition.
Another example of such a transition occurs in disordered high dimensional semiconductors
with dispersion relation $E_\mathbf{k}\sim |\mathbf{k}|^{\alpha^{\prime}}$ near a band edge.\cite{Syzranov:2015,Syzranov:2015c}
}
In this case the states near the bottom of the band get renormalized in the presence
of uncorrelated random potential for $d > 2\alpha'$. The average DOS vanishes at the critical point according to
Eqs.~(\ref{eq:rho-12}) and (\ref{eq:rho-13}) where to first order
in $\varepsilon'=d-2\alpha'$, the exponents read
\begin{eqnarray}
\nu^{\mathrm{Sem}} = \frac1{\varepsilon'}, \  \  \  \  \  z^{\mathrm{Sem}}= \alpha' + \frac{\varepsilon'}4. \label{eq-un-1}
\end{eqnarray}
Here we adapt the notation of Refs.~\onlinecite{Syzranov:2015,Syzranov:2015c} by putting a prime
to distinguish from the symbols already used in the present work.
The critical wave functions exhibit multifractality with the anomalous dimension given to one-loop order by~\cite{Syzranov2016}
\begin{eqnarray}
	\tilde{\Delta}_q^{\mathrm{Sem}} = \dfrac{1}{2}q(1-q) \varepsilon'					 +\mathcal{O}(\varepsilon^{\prime 2}).
\end{eqnarray}
The singularity spectrum peak is then located at
\begin{eqnarray}
	\alpha_0^{\mathrm{Sem}} = 2\alpha' + \dfrac{3}{2} \varepsilon'					 +\mathcal{O}(\varepsilon^{\prime 2}).
\end{eqnarray}
Using Eqs.~(\ref{eq:rho-13}) and~\eqref{eq-scal-rel-2} we find that the average and typical DOS vanish as (\ref{eq:rho-12}) and \eqref{eq-beta-typ-1} with the exponents given to first order by
\begin{eqnarray}
\beta^{\mathrm{Sem}} = \dfrac{3}{4} + \frac{\alpha' }{\varepsilon'}, \ \ \ \
	\beta_{\mathrm{typ}}^{\mathrm{Sem}} = \dfrac{5}{4} + \frac{\alpha' }{\varepsilon'}.  \label{eq-un-2}
\end{eqnarray}
For the  conventional case $\alpha'=2$, criticality is observed only in higher dimension, e.g. in $d=5$ which can be modeled numerically  using a tight-binding model on a lattice or  simulated using kicked quantum rotors.\cite{Syzranov:2015c}  In this situation $\varepsilon'=1$ and Eqs.~(\ref{eq-un-1}) and (\ref{eq-un-2})
give  $\nu^{\mathrm{Sem}}= 1$,   $z^{\mathrm{Sem}}=9/4$, $\beta^{\mathrm{Sem}} = 11/4$ and  $\beta_{\mathrm{typ}}^{\mathrm{Sem}}= 13/4$.
{\red Let us now turn back to the Weyl semimetal-diffusive metal transition.}

\section{\label{sec:RGflow} Renormalization group picture}

We now use a renormalization group (RG) approach to derive the multifractal spectrum, which is necessary to obtain $\alpha_0$, by computing the scaling dimension of a suitable composite operator for the disorder averaged theory. Let us first recall how to calculate the beta functions for the disorder strengths $\Delta_S$ (short-range correlated) and $\Delta_L$ (long-range correlated) following Ref.~\onlinecite{Louvet:2017}. We define the renormalized action as
\begin{multline}
\label{eq:S_renormalised}
S_R[\bar{\psi}_{\alpha},\psi_{\alpha}] =  \int_{k} \bar{\psi}_{\alpha}  (Z_{\psi} \bm{\gamma} \! \cdot \! \bm{k} - i Z_{\omega}\omega)
\psi_{\alpha} \\
- \frac{\mu^{-\varepsilon} Z_{S} \Delta_{S}}{K_d} \int_{k_i} (\bar{\psi}_{\alpha} \psi_{\alpha})(\bar{\psi}_{\beta} \psi_{\beta})\\
- \frac{\mu^{-\delta} Z_{L} \Delta_{L}}{K_d} \int_{k_i} k^{a-d} (\bar{\psi}_{\alpha} \psi_{\alpha})(\bar{\psi}_{\beta} \psi_{\beta}),
\end{multline}
where $K_d = 2/(4\pi)^{d/2} \Gamma(d/2)$ and $\mu$ is the mass scale at which we renormalize the theory. We use dimensional regularization to compute the renormalization $Z$ factors, which are introduced to render
all correlation functions finite.
Here we adopt the double expansion in $\varepsilon=d-2$ and $\delta=2-a$ developed in
Refs.~\onlinecite{Weinrib1983,Fedorenko:2006,Dudka:2016}. The relation between bare and
renormalized variables is given by
\begin{align}
\label{eq:field}
\mathring{\psi} &= Z_{\psi}^{1/2} \psi, &   \mathring{\omega} &= Z_{\omega}Z_{\psi}^{-1} \omega, \\
\label{eq:couplings}
\mathring{\Delta}_{S} &= \frac{2 \mu^{-\varepsilon}}{K_d}\frac{Z_{S}}{Z_{\psi}^2} \Delta_{S}, &  \mathring{\Delta}_{L} &= \frac{2 \mu^{-\delta}}{K_d}\frac{Z_{L}}{Z_{\psi}^2} \Delta_{L}.
\end{align}
where the upper circle denotes the bare quantity.

\begin{figure}
\includegraphics[scale=1]{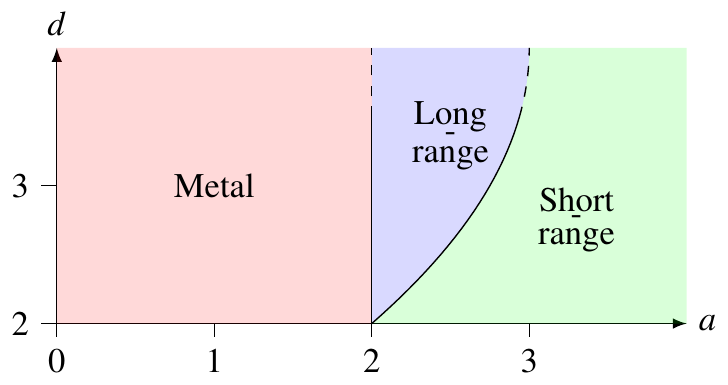}
\caption{Stability regions of different FPs in the plane $(a=2+\delta,d=2+\varepsilon)$.\cite{Louvet:2017}
For $a<2$ the Gaussian FP is unstable and the RG flow exhibits runaway reflecting instability of the
semimetal phase in the case of very LR correlated disorder:
a diffusive metal phase is settled for arbitrary weak disorder and the transition disappears.
The line $a_c(d) = d - (d-2)^2/4 + \mathcal{O}((d-2)^3)$
separates the regions where the critical behavior is controlled by the LR FP and SR FP.}
\label{fig:diagram_existence}
\end{figure}

The renormalization factors $Z_{S}$, $Z_{L}$, $Z_{\omega}$, and $Z_{\psi}$ have been computed to two-loop order in Ref.~\onlinecite{Louvet:2017}. The beta functions are defined as
\begin{equation}
\label{eq:beta_function}
\beta_i(\Delta_S,\Delta_L) =  - \mu \left. \dfrac{\partial \Delta_i}{\partial \mu} \right|_{\mathring{\Delta}_S,\mathring{\Delta}_L}, \ \ \ \ i=S,L
\end{equation}
and to two-loop order read
\begin{subequations}
\label{eq:beta-fun-all}
\begin{multline}
\beta_S =  - \varepsilon \Delta_S + 4 \Delta_S^2 + 4 \Delta_S \Delta_L \\ + 8 \Delta_S^3 + 20 \Delta_S^2 \Delta_L + 4 \Delta_L^3 + 16 \Delta_S \Delta_L^2, \label{eq:beta-fun1}
\end{multline}\vspace{-0.6cm}
\begin{multline}
\beta_L =  - \delta \Delta_L + 4 \Delta_L^2 + 4 \Delta_S \Delta_L \\ + 4 \Delta_L^3 + 4 \Delta_S^2 \Delta_L + 8 \Delta_S \Delta_L^2. \label{eq:beta-fun2}
\end{multline}
\end{subequations}
The beta functions (\ref{eq:beta-fun-all}) possess three fixed points (FPs) whose stability depends on the values of $\varepsilon$ and~$\delta$. The stability regions of these FPs are summarized in Fig.~\mbox{\ref{fig:diagram_existence}}.

\begin{enumerate}[label=(\roman*)]

	\item The Gaussian FP has $\Delta_S^G = \Delta_L^G = 0$, and its basin of attraction in the plane $(\Delta_S,\Delta_L)$ at fixed $\varepsilon$ and $\delta$
 gives the semimetal phase. It is unstable for $\delta<0$, which means the semimetal phase is unstable
 for very long-range correlated disorder.

\item The short-range fixed point (SR FP)
	\begin{align}
	\Delta_S^{{\rm SR}} & = \dfrac{\varepsilon}{4} - 	\dfrac{\varepsilon^2}{8} + \mathcal{O}	 \label{eq:SRFP1}		 (\varepsilon^3),\\
	\Delta_L^{{\rm SR}} & = 0, \label{eq:SRFP2}
	\end{align}
	 has a single unstable direction for
   $\delta > \delta_c = \varepsilon - \varepsilon^2/4 + \mathcal{O}(\varepsilon^3)$ and thus describes the transition leading to the same universality class as in the case of uncorrelated disorder.
   The critical exponents to two-loop are
   $1/\nu^{\mathrm{SR}} =\varepsilon + {\varepsilon^2}/{2} + O(\varepsilon^3)$ and
  $z^{\mathrm{SR}} = 1+ {\varepsilon}/{2} -{\varepsilon^2}/{8} + O(\varepsilon^3)$.

	\item The long-range fixed point (LR FP)
	\begin{align}
	\Delta_S^{\mathrm{LR}} & = \dfrac{\delta^3}{16(\varepsilon-\delta)} + \mathcal{O}	 \label{eq:LRFP1}		 (\varepsilon^3,\delta^3),\\
	\Delta_L^{\mathrm{LR}} & = \dfrac{\delta}{4} - \dfrac{\delta^2\varepsilon}{16(\varepsilon-\delta)} + \mathcal{O}			 (\varepsilon^3,\delta^3),  \label{eq:LRFP2}
	\end{align}
	 has a single unstable direction for $0 < \delta < \delta_c$ where it leads
  to a new universality class with
  $1/\nu^{\mathrm{LR}} = \delta +{\delta ^2 (2 \delta +\varepsilon )}/ {4 \varepsilon } + O(\varepsilon^3, \delta^3)$ and $z^{\mathrm{LR}} = 1+ {\delta}/{2}$ which is argued to be exact.

  \end{enumerate}

\begin{figure}
\includegraphics[width=\columnwidth]{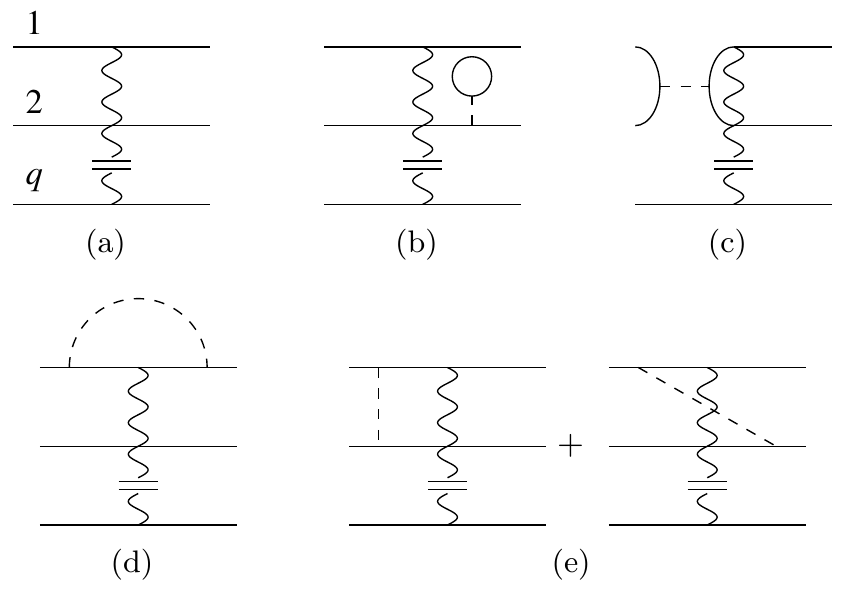}
\caption{\label{fig:figure2} {\red We use the graphical conventions of Ref.~\onlinecite{Syzranov2016},
with the difference that here each dashed line can represent either SR or LR disorder vertices.}
(a) Bare vertex $\mathring{\mathcal{O}}_q$. The horizontal solid lines stand for $\bar\psi_\alpha \psi_\alpha$, $\alpha =1,...,q$;
diagram (b) carries a loop which vanishes in the limit $n \rightarrow 0$;
diagram (c) in which two lines of different replicas are connected by a propagator is forbidden by definition~(\ref{eq:operator});
 diagram (d) has a combinatorial factor of $2q$ and its contribution is $\Delta_S/\varepsilon + \Delta_L/\delta$;
 diagrams (e) cancel each other.  }
\end{figure}

\begin{figure*}
\begin{center}
\includegraphics[scale=1]{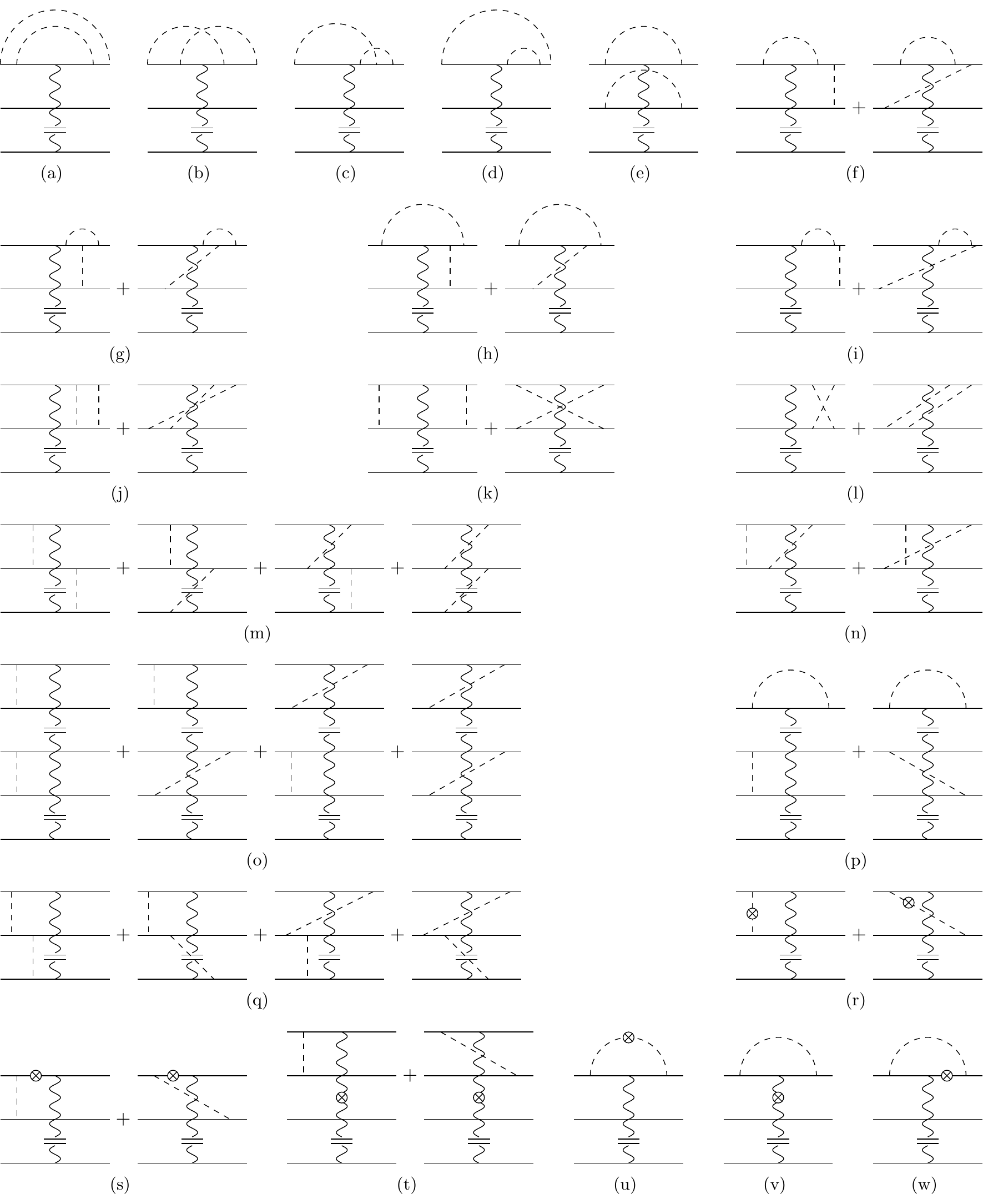}
\caption{(a) -(q) one-particle irreducible diagrams contributing to ${\Gamma}^{(0)}_{\mathcal{O}_q}$ to two-loop order. {\red These diagrams are topologically identical to those considered in Ref.~\onlinecite{Syzranov2016}, but here each dashed line corresponds to either SR or LR disorder vertices ($\Delta_S$ or $\Delta_L$), which makes the computation of these diagrams drastically complicated;}
 (r)-(w) counterterm diagrams. The cross represents the one-loop correction to the quartic interaction (dashed line), the propagator (solid line) or the operator $\mathcal{O}_q$ (wavy line). }
\label{fig:2loop_scalar}
\end{center}
\end{figure*}

We now show how to compute the multifractal spectrum within this framework. The replica trick enables to construct a proper composite operator whose scaling dimension corresponds to the moments of the local DOS,\cite{Foster:2012}
\begin{equation}
\label{eq:operator}
\mathcal{O}_q(\bm{r}) = \prod\limits_{\alpha=1}^q |\psi_{\alpha}(\bm{r})|^2,
\end{equation}
where $\alpha$ stands for the replica index and the product in Eq.~(\ref{eq:operator}) is taken  over $q$ distinct replicas. The scaling dimension  $x^*_q$ of the operator $\mathcal{O}_q$ can be straightforwardly computed
from the renormalization constant $Z_q$, defined as
\begin{equation}
\mathring{\mathcal{O}}_q = Z_q Z_{\psi}^{-q} \mathcal{O}_q. \label{eq-Zq-ren0}
\end{equation}
Renormalization condition~(\ref{eq-Zq-ren0}) renders the renormalized vertex functions with
insertion of a single composite operator~(\ref{eq:operator})  to be finite,
\begin{equation}
\mathring{\Gamma}^{(\mathcal{N})}_{\mathcal{O}_q}(\{\bm{r}\};\mathring{\omega}, \mathring{\Delta}) = Z_q Z_{\psi}^{-\frac{\mathcal{N}}2-q} \Gamma^{(\mathcal{N})}_{\mathcal{O}_q}(\{\bm{r}\};\omega,\Delta,\mu), \label{eq-Zq-ren}
\end{equation}
where $\mathcal{N}$ is the number of external legs $\bar{\psi},\psi$.
The renormalization constant $Z_q$ can be found from renormalization of the vertex function $\mathring{\Gamma}^{(0)}_{\mathcal{O}_q}$.
The one- and two-loop  diagrams contributing to this vertex function
are shown in Figs.~\ref{fig:figure2} and \ref{fig:2loop_scalar}, respectively.
The corresponding values of diagrams with combinatorial factors are summarized in
Tab.~\mbox{\ref{tab:2loop_scalar}}.

\begin{table*}
\begin{center}
\renewcommand{\arraystretch}{1.8}
\begin{tabular*}{\linewidth}{@{\extracolsep{\fill}}ccccc}
\hline
Diagram & Combinatorial factor & $\Delta_S^2$ & $\Delta_S \Delta_L$ & $\Delta_L^2$ \\
\hline
\hline
$(a)$ & $4q$ & $\frac{1}{\varepsilon^2} + \frac{2}{\varepsilon}$ & $\frac{1}{\varepsilon \delta} + \frac{2}{\delta(\delta+\varepsilon)} + \frac{3}{\delta}  + \frac{1}{\varepsilon}$ & $\frac{3\delta-\varepsilon}{\delta(2\delta-\varepsilon)} + \frac{3\delta - \varepsilon}{2 \delta^2(2 \delta-\varepsilon)}$ \\
$(b)$ & $4q$ & $-\frac{1}{2\varepsilon^2}-\frac{1}{\varepsilon}$ & $-\frac{2}{\delta(\delta+\varepsilon)} - \frac{2}{\delta}$ & $-\frac{1}{2\delta(2\delta-\varepsilon)} -\frac{1}{2\delta-\varepsilon}$ \\
$(c)$ & $8q$ & $\frac{1}{2\varepsilon^2}+\frac{1}{\varepsilon}$ & $\frac{2}{\delta(\varepsilon+\delta)} +  \frac{1}{\varepsilon} + \frac{1}{\delta}$ & $\frac{1}{2\delta^2} + \frac{1}{\delta}$ \\
$(d)$ & $8q$ & $\frac{1}{2\varepsilon}$ & $\frac{1}{2\varepsilon}+\frac{1}{2\delta} - \frac{\delta -\varepsilon}{2\delta(\varepsilon + \delta)}$ & $\frac{1}{2 \delta} - \frac{\delta-\varepsilon}{4 \delta^2}$ \\
$(e)$ & $2q(q-1)$ & $ \frac{1}{\varepsilon^2} + \frac{2}{\varepsilon}$ & $ \frac{2}{\varepsilon \delta} + \frac{2}{\varepsilon} + \frac{2}{\delta}$ & $\frac{1}{\delta^2} + \frac{2}{\delta}$ \\
$(f)$ & $8q(q-1)$ & $\frac{1}{\varepsilon}$ & $\frac{1}{\varepsilon} + \frac{1}{\delta}$ & $\frac{1}{\delta}$ \\
$(g)$ & $8q(q-1)$ & $\frac{1}{\varepsilon}$ & $\frac{1}{\varepsilon} + \frac{1}{\delta}$ & $\frac{1}{\delta}$ \\
$(j)$ & $4q(q-1)$ & $\frac{1}{\varepsilon^2} + \frac{1}{2\varepsilon}$ & $\frac{1}{\varepsilon \delta} + \frac{2}{\delta(\delta+\varepsilon)} + \frac{1}{2\varepsilon} + \frac{1}{\delta+\varepsilon}$ & $\frac{3 \delta - \varepsilon}{2\delta^2(2\delta-\varepsilon)} + \frac{3\delta-\varepsilon}{4\delta^2}$\\
$(k)$ & $2q(q-1)$ & $\frac{1}{\varepsilon^2} + \frac{1}{2\varepsilon}$ & $\frac{1}{\varepsilon \delta} + \frac{2}{\delta(\delta+\varepsilon)} + \frac{1}{2\varepsilon} + \frac{1}{\delta+\varepsilon}$ & $\frac{3 \delta - \varepsilon}{2\delta^2(2\delta-\varepsilon)} + \frac{3\delta-\varepsilon}{4\delta^2}$\\
$(l)$ & $4q(q-1)$ & $-\frac{1}{2\varepsilon^2} - \frac{1}{2\varepsilon}$ & $-\frac{2}{\delta(\delta+\varepsilon)} - \frac{2}{\varepsilon + \delta}$ & $-\frac{1}{2\delta(2\delta-\varepsilon)} - \frac{1}{2\delta}$\\
$(n)$ & $8q(q-1)$ & $-\frac{1}{2\varepsilon^2} - \frac{1}{2\varepsilon}$ & $-\frac{2}{\delta(\delta+\varepsilon)} - \frac{2}{\varepsilon + \delta}$ & $-\frac{1}{2\delta(2\delta-\varepsilon)} - \frac{1}{2\delta}$\\
$(p)$ & $4q(q-1)(q-2)$ & $\frac{1}{\varepsilon}$ & $\frac{1}{\varepsilon} + \frac{1}{\delta}$ & $\frac{1}{\delta}$\\
$(r)$ & $2q(q-1)$ & $-\frac{4}{\varepsilon}$ & $-\frac{4}{\varepsilon}-\frac{4}{\delta}$ & $-\frac{4}{\delta}$\\
$(t)$ & $2q(q-1)$ & $-\frac{2q}{\varepsilon}$ & $-\frac{2q}{\varepsilon} -\frac{2q}{\delta}$ & $-\frac{2q}{\delta}$\\
$(u)$ & $2q$ & $-\frac{4}{\varepsilon^2}- \frac{4}{\varepsilon}$ & $-\frac{8}{\varepsilon \delta} - \frac{4}{\varepsilon} - \frac{4}{\delta}$ & $-\frac{4}{\delta^2}- \frac{4}{\delta}$\\
$(v)$ & $2q$ & $-\frac{2q}{\varepsilon^2}- \frac{2q}{\varepsilon}$ & $-\frac{4q}{\varepsilon \delta} - \frac{2q}{\varepsilon} - \frac{2q}{\delta}$ & $-\frac{2q}{\delta^2}- \frac{2q}{\delta}$\\
$(w)$ & $4q$ & $-\frac{1}{\varepsilon}$ & $-\frac{1}{\varepsilon} - \frac{1}{\delta}$ & $-\frac{1}{\delta}$ \\
\hline
\end{tabular*}
\caption{ Poles of diagrams depicted in Fig.~\ref{fig:2loop_scalar}. Diagrams (h), (i), (m), (o), (q) and (s) cancel each other. The last three columns give the terms proportional to $\Delta_S^2$, $\Delta_S \Delta_L$ and $\Delta_L^2$, respectively. In computing these diagrams one encounters many different types of integrals which can be found in Appendix B of Ref.~\onlinecite{Dudka:2016}. We express the vertices, propagators and composite operators in terms of the renormalized parameters ($\Delta_S$, $\Delta_L$, $\omega$ and $\mathcal{O}_q$) instead of the bare parameters ($\mathring{\Delta}_S$, $\mathring{\Delta}_L$, $\mathring{\omega}$ and $\mathring{\mathcal{O}}_q$) and to compensate this reparametrization we add the counterterms (diagrams (r) -- (w)).
}
\label{tab:2loop_scalar}
\renewcommand{\arraystretch}{1}
\end{center}
\end{table*}


We can now write down the RG flow equation for the $q$-moment of the local DOS:
\begin{eqnarray}
\left[\sum\limits_{i=S,L} \beta_i(\Delta) \frac{\partial}{\partial \Delta_i} + z (\Delta) \omega \dfrac{\partial}{\partial \omega}
 -x_q(\Delta) \right] \overline{\rho^q} (\Delta,\omega) = 0, \label{eq:callan_symanzik} \nonumber \\
\end{eqnarray}
where the $\beta$ functions are given by Eqs.~(\ref{eq:beta-fun-all}) and
\begin{subequations}
\begin{eqnarray}
&&z (\Delta)  = 1+ \eta_{\omega}(\Delta) - \eta_{\psi} (\Delta), \\
&&x_q(\Delta) = (d-1+\eta_\psi)q-\eta_q, \label{eq:xq-def} \\
&& \eta_{j}(\Delta) = - \sum\limits_{i =S,L} \beta_i \dfrac{\partial \ln Z_{j}}{\partial \Delta_i}, \label{eq:eta-def}\ \ \
j = \psi,\omega,q.
\end{eqnarray}
\end{subequations}
We can solve Eq.~(\ref{eq:callan_symanzik}) using the method of characteristics. In the vicinity of a FP $\Delta^*=(\Delta^*_S,\Delta^*_L)$ with one unstable direction we find that
\begin{eqnarray}
 \overline{\rho^q} (\Delta,\omega) = \xi^{-x_q^*} f (\omega \xi^z, |\Delta -\Delta^*| \xi^{1/\nu}), \label{eq:rho-q-sol}
\end{eqnarray}
where $x_q^*=x_q(\Delta^*)$ and $\xi$ is the correlation length. Using Eq.~(\ref{eq:eta-def}) we find
\begin{subequations}
\begin{eqnarray}
 \eta_{\psi}(\Delta) &=& -2 \Delta_S^2+2 \Delta_L^2  -\frac{4  \varepsilon }{\delta }\Delta_L (\Delta_S + \Delta_L),   \label{eq:eta-psi-exp} \\
 \eta_q (\Delta)  &=& q \left[ 2 (\Delta_S + \Delta_L) - 6 \Delta_S^2 + \left(1 - \frac{7\varepsilon}{\delta} \right) \Delta_L^2 \right. \nonumber \\ \nonumber
&& - \left. \Delta_S \Delta_L \left(11 + \frac{4\varepsilon}{\delta} - \frac{3 \delta}{\varepsilon} \right) \right] + q^2 \left[ 6 \Delta_S^2 \right. \\
&& \left.  + 3 \left(5 - \frac{\delta}{\varepsilon} \right) \Delta_S \Delta_L \right.
 + \left. 3 \left(1 + \frac{\varepsilon}{\delta} \right) \Delta_L^2 \right]\!,~~~\\
 \eta_\omega (\Delta) &=& \eta_{q=1}(\Delta).
\end{eqnarray}
\end{subequations}
From the last equation we recover for $q=1$ the dynamic critical exponent $x^*_{q=1}=d-z$, as expected.
Using  Eqs.~(\ref{eq:tau-def}) and (\ref{eq:xq-def})  we obtain
\begin{multline}
\tilde{\Delta}_q = q(1-q)\left[ 6 \Delta_S^2 + 3 \left(5 - \frac{\delta}{\varepsilon} \right) \Delta_S \Delta_L \right. \\
\left. + 3  \left(1 + \frac{\varepsilon}{\delta} \right) \Delta_L^2 \right]. \label{eq:Deltaqq}
\end{multline}
To compute the critical anomalous dimension we have to evaluate~(\ref{eq:Deltaqq}) at the corresponding FP  $\Delta^*$.

\begin{enumerate}[label=(\roman*)]
	\item For $\delta > \delta_c$ the critical behavior is controlled by the SR FP. Substituting (\ref{eq:SRFP1})-(\ref{eq:SRFP2})  into (\ref{eq:Deltaqq}) we recover the result of Refs.~\onlinecite{Louvet:2016,Syzranov2016},
	\begin{equation}
	\label{eq:spectrum_SR}
	\tilde{\Delta}_q^{\mathrm{SR}} = \dfrac{3}{8}q(1-q) 			\varepsilon^2 +\mathcal{O}(\varepsilon^3).
	\end{equation}
	\item For $0 < \delta < \delta_c$ the critical behavior is controlled by the LR FP. Substituting (\ref{eq:LRFP1})-(\ref{eq:LRFP2})  into (\ref{eq:Deltaqq}) we find the anomalous dimension corresponding to the new universality class
	\begin{equation}
	\label{eq:spectrum_LR}
	\tilde{\Delta}_q^{\mathrm{LR}} = \dfrac{3}{16}q(1-q) 		\delta(\delta + \varepsilon) 					 +\mathcal{O}(\varepsilon^3,\delta^3).
	\end{equation}
\end{enumerate}
It is easy to check that both results match on the line $\delta_c=\varepsilon + O(\varepsilon^2)$ which separates the two regions of stability.  For $\delta=0$, $\tilde{\Delta}_q$ vanishes, which is consistent with the disappearance of the transition.

The singularity spectrum~(\ref{eq:tau-def2}) corresponding to the multifractal spectra~(\ref{eq:spectrum_SR}) and (\ref{eq:spectrum_LR}) is quadratic, which implies a log-normal distribution   $\mathcal{P}(\tilde{\rho},L)$ for the local DOS. It can be expressed as
\begin{equation}
\tilde{f}(\alpha) = d- \frac{(\alpha-\alpha_0)^2}{4(\alpha_0-d)},
\end{equation}
which has a maximum at $\alpha=\alpha_0$.
\begin{enumerate}[label=(\roman*)]
\item For SR correlated disorder ($\delta > \delta_c$), we find
\begin{equation}\label{eq-sing-peak}
\alpha_0^{\mathrm{SR}} = 2 + \varepsilon + \dfrac{3}{8} \varepsilon^2 +\mathcal{O}(\varepsilon^3).
\end{equation}
The $[1/1]$ Pad\'{e} approximant of Eq.~(\ref{eq-sing-peak}) gives $\alpha_0^{\mathrm{SR}} =3.6$ in three dimensions, in fair agreement  with the  numerical prediction of Eq.~\eqref{eq:alpha-zero-num}.
\item For LR correlated disorder ($0 <\delta < \delta_c$), we find
\begin{equation}\label{eq-sing-peak2}
\alpha_0^{\mathrm{LR}}  = 2 + \varepsilon + \dfrac{3}{16} \delta(\varepsilon +\delta)+\mathcal{O}(\varepsilon^3,\delta^3).
\end{equation}
In this case $\alpha_0$ is smaller and thus the distribution of local DOS is thinner.
\end{enumerate}
One can compare these results with that for the Anderson localization transition.
In  the three dimensional orthogonal class one finds $\alpha_0 = 4$ to two-loop order, wich is in  excellent agreement with the numerical result $\alpha_0 = 4.03 \pm 0.05$.\cite{Mildenberger2002} Thus multifractality is stronger at the Anderson localization than at the SR semimetal-metal transition, which is itself stronger than at the LR semimetal-metal transition.

{\red
Let us summarize our main findings. We have derived new scaling relations  \eqref{eq-scal-rel-2} and \eqref{beta-typ-scaling} which presumably hold not only for the Weyl semimetal-diffusive metal transition, but for all disorder-driven transitions wherein the typical and average local DOS vanish on one side of the critical point, as discussed in Sec.~\ref{sec:unconventional}. We have computed the multifractal spectra (\ref{eq:spectrum_SR}) and (\ref{eq:spectrum_LR}) of the critical wave functions at the semimetal-diffusive metal transition for SR and LR correlations of disorder. This enabled us to find the position of the peak in the singularity spectrum, $\alpha_0$, which is consistent with previous numerical simulations with SR correlated disorder. To characterize completely the SR and LR universality classes, we now show that vector potential disorder is an irrelevant perturbation.}

\section{\label{sec:vectorial}
Vector potential disorder}

Uncorrelated vector potential disorder is known to have no effect on criticality at the semimetal-diffusive metal transition.\cite{Sbierski:2016}
In this section we demonstrate the irrelevance of vector potential disorder even in the presence of LR disorder correlations, unless it is so long-range correlated ($a<2$) that it destabilizes the semimetal phase.
Since the time-reversal symmetry of the Weyl Hamiltonian is accidental it is natural
to include a general disorder potential that breaks time-reversal invariance,
\begin{equation}
V(\bm{r}) = \sum_{\mu=0}^3 V_{\mu}(\bm{r}) \sigma_{\mu},
\end{equation}
were $\sigma_0 = \mathbb{I}$ is the identity matrix, $\sigma_i$ with $i=1,2,3$ are the Pauli matrices, $V_0(\bm{r})$ is a scalar potential  and $V_i(\bm{r})$ is a random vector potential.
We assume the absence of mutual correlations between different components of disorder potential and that
the strength of disorder is isotropic, i.e.
\begin{equation}
 \overline{V_{\mu}(\bm{r}) V_{\mu'}(\bm{0})} =  g_{\mu}(r) \delta_{\mu\mu'}
\end{equation}
for $\mu = 0,...,3$, with $g_1(r) = g_2(r) = g_3(r)$.

In the case of vector potential disorder, dimensional regularization leads to the appearance of
evanescent operators already at one loop order.\cite{Louvet:2016} To avoid this problem we adopt here a different regularization scheme based on the so-called  $\varepsilon_m$-expansion (see Ref.~\onlinecite{Roy2016} for further details). In this scheme we work in fixed dimension $d=3$ and regularize the effective action in the ultraviolet by setting
\begin{align}
\label{eq:correlation_vect}
g_0(k) = \Delta k^{-m}, && g_i(k) = \kappa k^{-l}
\end{align}
for $i=1,2,3$, and expand in the small parameters $\varepsilon_m=1-m$ and $\varepsilon_l=1-l$. This scheme has the advantage to preserve a finite Clifford  algebra of $\gamma$ matrices~\cite{Kennedy1981} and to include naturally long-range correlations, with independent and tunable parameters $a_m=2+\varepsilon_m$  and $a_l=2+\varepsilon_l$ for scalar and vector potential disorder, respectively. We can study the short-range correlations simply by choosing $\varepsilon_m = 1$ or $\varepsilon_l = 1$. The renormalized action now reads
\begin{multline}
\mathcal{S}_R[\bar{\psi}_{\alpha}, \psi_{\alpha}] = \int_{k,\omega} \bar{\psi}_{\alpha}  (Z_{\psi} \bm{\gamma} \! \cdot \! \bm{k} - i Z_{\omega}\omega)
\psi_{\alpha}\\
- \frac{\mu^{-\varepsilon_m} Z_{\Delta} \Delta}{K_d} \int_{k_i,\omega_i} k^{-m} (\bar{\psi}_{\alpha} \psi_{\alpha})( \bar{\psi}_{\beta} \psi_{\beta})\\
- \frac{\mu^{-\varepsilon_l} Z_{\kappa} \kappa}{K_d} \sum_{i=1}^3 \int_{k_i,\omega_i} k^{-l} (\bar{\psi}_{\alpha}\sigma_{i}\psi_{\alpha})( \bar{\psi}_{\beta} \sigma_{i} \psi_{\beta}).
\end{multline}
The relations between bare and renormalized parameters are similar to Eqs.~(\mbox{\ref{eq:field}}) and (\mbox{\ref{eq:couplings}}):
\begin{align}
\mathring{\psi} = Z_{\psi}^{1/2} \psi,    && \mathring{\omega} = Z_{\omega}Z_{\psi}^{-1} \omega,
\end{align}
\begin{align}
\mathring{\Delta} = \frac{2 \mu^{-\varepsilon_m}}{K_d}\frac{Z_{\Delta}}{Z_{\psi}^2} \Delta, &&
\mathring{\kappa} = \frac{2 \mu^{-\varepsilon_l}}{K_d}\frac{Z_{\kappa}}{Z_{\psi}^2} \kappa.
\end{align}
We  compute the renormalization constants $Z_{\Delta}$, $Z_{\kappa}$, $Z_{\psi}$ and $Z_{\omega}$  in the minimal subtraction scheme to one-loop order.
\begin{figure}
\begin{center}
\includegraphics[scale=1]{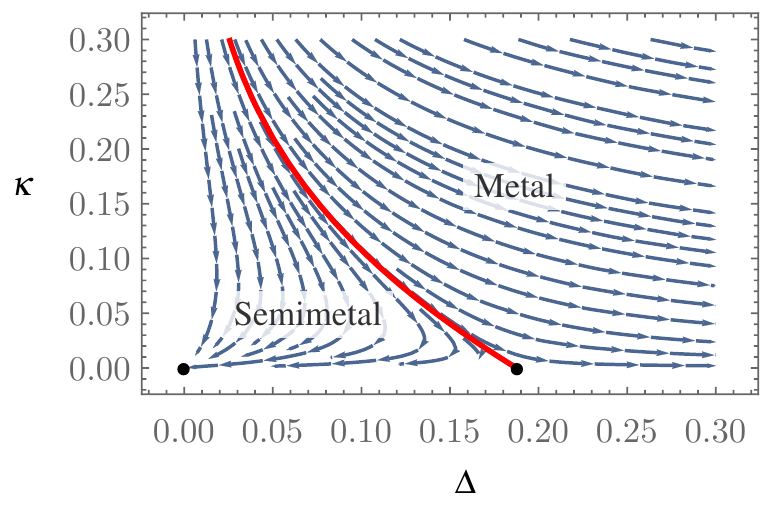}
\caption{Renormalization flow and phase diagram for $\varepsilon_m=1$ and $\varepsilon_l=1$ (short-range disorder). The black dots are the Gaussian and non-trivial (scalar) fixed points. The red thick line is the separatrix between the Gaussian basin of attraction (semimetal phase) and the runaway behavior (metal phase). For small $\Delta$ its asymptotics is given by  $\kappa \approx -3\varepsilon_l \ln (\Delta) /32$.}
\label{fig:vectorial_flow}
\end{center}
\end{figure}
\begin{equation}
Z_{\Delta} = 1 + \dfrac{4 \Delta}{\varepsilon_m} + \dfrac{12 \kappa}{\varepsilon_l} + \mathcal{O}(\Delta^2, \kappa^2),
\end{equation}
\begin{equation}
Z_{\kappa} = 1 - \dfrac{4 \Delta}{3 \varepsilon_m} + \dfrac{4 \kappa}{3 \varepsilon_l} + \mathcal{O}(\Delta^2, \kappa^2).
\end{equation}
\begin{equation}
Z_{\psi} = 1 - \dfrac{2 \Delta}{3\varepsilon_m}+ \dfrac{2 \kappa}{3\varepsilon_l} + \mathcal{O}(\Delta^2, \kappa^2)
\end{equation}
\begin{equation}
Z_{\omega} = 1 + \dfrac{2 \Delta}{\varepsilon_m} + \dfrac{6 \kappa}{\varepsilon_l} + \mathcal{O}(\Delta^2, \kappa^2).
\end{equation}
The beta functions are defined as in Eq.~\mbox{\eqref{eq:beta_function}},
\begin{align}
\beta_\Delta =  - \mu \left. \dfrac{\partial \Delta}{\partial \mu} \right|_{\mathring{\Delta}, \mathring{\kappa}}, && \beta_\kappa =  - \mu \left. \dfrac{\partial \kappa}{\partial \mu} \right|_{\mathring{\Delta}, \mathring{\kappa}},
\end{align}
and have the following expressions,
\begin{align}
\beta_\Delta & = - \varepsilon_m \Delta + \dfrac{16}{3} \Delta^2 + \dfrac{32}{3} \Delta \kappa, \\
\beta_\kappa & = - \varepsilon_l \kappa.
\end{align}
Notice that the one-loop terms of $Z_\psi$ and $Z_\kappa$ cancel out in the beta function $\beta_\kappa$. These flows are consistent with those found in Dirac semimetals when only chiral preserving disorder is allowed, \cite{Roy2016} and with previous studies of disordered Weyl nodes using the Wilson  renormalization  scheme.\cite{Sbierski:2016} Figure~\mbox{\ref{fig:vectorial_flow}} shows the renormalization flow in the case of SR correlated disorder ($\varepsilon_m=1$ and $\varepsilon_l=1$).
Apart from the trivial Gaussian fixed point, the only nontrivial FP is
\begin{align}
\Delta^* =\dfrac{3\varepsilon_m}{16}, && \kappa^* =0. \label{eq-new-FP-vector}
\end{align}
The corresponding stability matrix has eigenvalues $\varepsilon_m$ and $-\varepsilon_l$ so that the FP~(\ref{eq-new-FP-vector}) is relevant (and thus controls criticality) for $\varepsilon_l > 0$. This conclusion holds whether disorder is short-range ($\varepsilon_m = 1$ or $\varepsilon_l = 1$) or long-range ($0 < \varepsilon_m < 1$ or $0 < \varepsilon_l < 1$), but the region of stability for the semimetal phase shrinks with decreasing $\varepsilon_l$ until it disappears at $\varepsilon_l=0$. {\red Hence vector potential disorder does not affect criticality. We do not claim, however, that the presence of only vector potential disorder cannot induce a transition, which naively follows from the one-loop RG flow shown in Fig.~\ref{fig:vectorial_flow}. Indeed the separatrix in Fig.~\ref{fig:vectorial_flow} may hit the $\kappa$ axis if higher order terms were included in the beta functions, which would be consistent with the numerical simulations of Ref.~\onlinecite{Sbierski:2016}. }

\section{\label{sec:discussion} Summary}

We have studied the multifractality of critical wave functions at the Weyl semimetal-diffusive metal transition for
the most general disorder, including random scalar and vector potentials with both short-range and long-range correlations. Using a renormalization group method we have computed the multifractal spectrum to two-loop order as a function of the disorder correlation exponent $a$. The multifractal spectrum is an alternative way to characterize the transition, which is both richer and more accurate than the conventional critical exponents.

We have related the multifractal spectrum to the distribution of the local DOS fluctuations and studied
the behavior of the average and typical local DOS  near the critical point, which scale as power-laws with two different exponents  $\beta$ and $\beta_{\text{typ}}$ respectively.
We have derived the new scaling relation \eqref{eq-scal-rel-2}, which is in fair agreement with the known numerical results for uncorrelated disorder, and valid for other quantum disorder-driven phase transitions in which both the average and  typical local  DOS vanish on one side of the transition. In particular the relation holds for the unconventional quantum transition in disordered semiconductors with power-law dispersion relation near the band edge. We are confident that our findings will stimulate new numerical studies on multifractality and the effects of disorder correlations at the Weyl semimetal-diffusive metal transition and other disorder-driven quantum phase transitions.\\

\acknowledgments

We would like to thank I.~Balog,  V.~Juri\v{c}i\'{c}, B.~Roy and B.~Sbierski for valuable discussions.
We acknowledge  support from the French Agence Nationale de la Recherche  by the Grant No. ANR-17-CE30-0023 (DIRAC3D).

\bibliographystyle{apsrev4-1}

%

\end{document}